\documentclass{cfdsc}
\usepackage[T1]{fontenc}
\usepackage[latin1]{inputenc}
\usepackage{setspace}
\usepackage{amssymb}
\usepackage{amsmath}
\usepackage{float}
\usepackage{multirow}
\usepackage{tabularx}
\usepackage[flushleft]{threeparttable}
\usepackage{booktabs,caption,fixltx2e}
\usepackage{relsize}
\makeatletter

\usepackage{xstring}
\usepackage{tikz}
\usetikzlibrary{calc}

\newcommand{\mypm}{\mathbin{\mathpalette\@mypm\relax}}
\newcommand{\@mypm}[2]{\ooalign{%
  \raisebox{.1\height}{$#1+$}\cr
  \smash{\raisebox{-.6\height}{$#1-$}}\cr}}
\makeatother
\newcommand{\rpm}{\raisebox{.2ex}{$\scriptstyle\pm$}}

\def\NumOfColumns{8}%
\def\Sequence{1/A, 2/B, 3/C, 4/D, 5/E, 6/F, 7/G, 8/H}%

\newcommand{\Size}{0.75 cm}
\tikzset{Square/.style={
    inner sep=0pt,
    text width=\Size, 
    minimum size=\Size,
    draw=black,
    fill=white!20,
    align=center,
    }
}

\begin{document}

\title{Uncertainty quantification of tabulated supercritical thermodynamics for compressible Navier-Stokes solvers}

\author{Sai Praneeth}

\institute{Department of Mechanical and Mechatronics Engineering, University of Waterloo, Waterloo, Canada\\200 University Ave. W., Waterloo, Ontario, N2L 3G1}

\email{spmuppar@uwaterloo.ca}

\maketitle

\begin{abstract}
Non-ideal state equations are needed to compute a growing number of engineering-relevant problems. The additional computational overhead from the complex thermodynamics accounts for a significant portion of the total computation, especially the near-critical or transcritical thermodynamic regimes. A compromise between computational speed and the accuracy of the thermodynamic property evaluations results in a propagation of the error from the thermodynamics to the hydrodynamic computations.  This work proposes a systematic error quantification and computational cost estimate of the various approaches for equation of state computation for use in compressible Navier-Stokes solvers in the supercritical regime. We develop a parallelized, high-order, finite volume solver with highly-modular thermodynamic implementation to compute the compressible equations in conservative form. Three tabular approaches are investigated: homogeneous tabulation, block structured adaptive mesh refinement tabulation, and a n-dimensional Bezier surface patch on an adaptive structured mesh. We define a set of  standardized error metrics and evaluate the thermodynamic error, table size and computational expense for each approach. We also present an uncertainty quantification methodology for tabular equation of state.
\end{abstract}

\section{Introduction}
The simulation of thermodynamically complex systems present new challenges to the computational fluid dynamics community. Recent progress has been made in predictive modelling of trans- and supercritical flows to address design challenges in supercritical water-cooled reactors (SCWR) \cite{Cai2014}, understanding thermoacoustic oscillation in transcritical heat exchangers, efficient supercritical extraction \cite{Aizpurua-Olaizola2015}, cryogenic mixing and combustion in liquid rocket engines \cite{Hickey2013}, and technologies relating to carbon capturing \cite{Stucki2009}. These flows are characterized by strong thermodynamic non-linearities that arise, primarily, in the near-critical region of the fluid and require special considerations for efficient computational modelling. In its essence, the Navier-Stokes equations are a set of partial differential equations that govern the transport of the conservative properties of dynamic fluid system. Typically, the governing equations are written in terms of density, momentum and total energy. These governing equations must be closed using an equation of state (EOS) that is needed to compute the contribution due to local pressure and thermal gradients in the system. The EOS is also needed to obtain an estimate of the transport properties (e.g. viscosity, thermal conductivity), which depend on the local thermodynamic state.

For a single phase, single species fluid, two independent thermodynamic variables are needed to fully define the local thermodynamic state. The compressible Navier-Stokes equation fully define the thermodynamic state of the system as both the density ($\rho$) and internal energy ($e$) can be computed directly from the conservative variable matrix: $\left[\rho, \rho \vec{u}, \rho E\right]$, where $\rho E$ typically, represents the total energy (internal plus kinetic). Most CFD solvers require the state equation to transform the variables from conservative to their primitive state: $\left[\rho, \vec{u}, p, T\right]$.  Under a perfect gas assumption, we trivially find  $p=\rho (\gamma -1 ) e$ which gives us, using the EOS, the solution of $T=p/(\rho R)$; under an ideal gas assumption the complexity rises slightly with the non-constant definition of the specific heat. For real fluids, there is no explicit relation between $(T,p)$ and $(\rho, e)$ from classical analytical state equations. This typically requires an iterative algorithm to find the correct $(T,p)$ tuple for the known $(\rho, e)$ from the conservation equations. Given the complexity of the EOS in the supercritical regime\textemdash its simplest form being a highly non-linear cubic equation\textemdash this thermodynamic computation demands non-negligible computational resources. Furthermore, due to the inherent thermodynamic non-linearities about pseudo-boiling point, small errors in the thermodynamic property estimation can lead to large variation in the computed pressure and temperature and concomitantly directly affect the thermodynamic gradient calculations in the flow. This, in turn, may negatively  affect in the computed integral properties of the simulation. 

The added computational expense required for complex state equations quickly becomes a limiting factor in large scale computations. Typical real-fluid simulations with cubic state equations will show slow downs of 50\% to 100\% compared to ideal gas simulations depending on the thermodynamic state of the simulation.  Masquelet \cite{Masquelet2006} notes  that over 20\% of the computational expenses in the real fluid calculations comes from expensive numerical operations such as square-roots, logarithms and iterative algorithms.  The  overall computational expense rises drastically with more complex analytical state equation evaluations (e.g. Modified Benedict-Webb-Rubin). For this reason,  cubic class of state equations (e.g. Peng-Robinson-Stryjek-Vera, Soave-Redlich-Kwong etc.) have become the standard  when using analytical thermodynamic definitions for CFD, in spite of their sub-optimal representation of the actual thermodynamics.  

The predictive capability of fluid solvers is heavily dependent on the accuracy of the EOS in the transcritical regime. Given the limitations of a finite-degree polynomial-type function, typical cubic-state equations show a very significant errors compared to experimental values; errors of over 30\% are expected for the Peng-Robinson equation in the some parts of the state space \cite{Stryjek1986}.  Furthermore, within the subcritical, two-phase, the continuous state equation  has an inversion, meaning that the isobaric lines  do not behave monotonically which can lead to non-physical behavior and numerical instabilities. 

Therefore, a compromise needs to be struck between thermodynamic accuracy and computational efficiency. To more closely control the thermodynamic error and, in some cases, accelerate the computational cost related to the EOS computations, a growing number of  codes use pre-tabulated state equations based on well-defined REFPROP quantities.  The tabulation accuracy is generally proportional to the size of the table. On the downside, the size of the table  may hinder efficient caching and prefetching for memory which has a significant cost on the computational speed of a simulation. It is also particularly problematic for large scale, memory-limited computation on high-performance computers.  There is surprising lack of quantitative information relating the thermodynamic accuracy and computational expense specifically related to CFD solvers. Most of the comparisons discuss the errors related to state equations,  and do not tackle the specificity of the thermodynamic calculations of CFD solvers which require solutions for known density and internal energy. There are some works that address novel ways to tabulate the state equations for CFD solvers \cite{Collins2013,Liu2014,Liu2015} and other works that evaluate the uncertainty propagation of thermodynamic error \cite{Cinnella2011}. The present work quantifies different state equation implementations and proposes quantifiable metrics to evaluate the accuracy and computational expenses of each approach.  The focus of this work  lies in the trans- and supercritical regime in which there is a first-order continuity in the state  properties. Finally, this work will present the resulting implications of thermodynamic error on simulation results of well defined  CFD cases at transcritical thermodynamic conditions. The following section describes the main details of our numerical framework. Afterward, the various tabulated approaches  used in the current work are presented followed by a set of metrics to evaluate the thermodynamic accuracy and computational cost for all tabulated approaches. Finally, statistical study is undertaken to quantify the propagation of uncertainties from the thermo- to the hydronamic quantities for well defined numerical simulations (this last point will be completed in the final paper).

\section{Numerical Framework}
A modular finite-volume code, named ThermoCode, was developed for the study of complex thermodynamic fluid systems. The code, which solves the compressible Navier-Stokes equations in fully conservative form, rests on a high-order flux reconstruction at the faces which requires a structured grid arrangement. The code is formally of fourth-order accuracy. The flux at each face is computed as the sum of a left- and a right biased stencil at the face both of second-order accuracy. The code has been parallelized with efficient flux passing between the processors which allows for a limited number of ghost cells and communication overhead. Explicit time advancement is done with low-storage, fourth-order Runge-Kutta scheme. The purpose of a high fidelity code is to remove, as much as possible, numerical dissipation from the thermodynamic error evaluation. The coupling between the Navier-Stokes equations and the thermodynamics and transport property modules are designed to be extremely modular. The code contains a number of different analytical cubic state equations (Peng-Robinson, Suave-Redlich-Kwong) as well as a number of tabular approaches discussed in this paper.

\section{Tabulated State Equations}
State equation tabulation involves storing the thermodynamic and transport properties of the species in a library (e.g. binary file, .xml file), by doing so we move the evaluation of the EOS to a preprocessing step. Thus decoupling the EOS computation complexity from runtime performance. This approach allows for the use of accurate high-quality EOS or REFPROP reference data without major runtime penalty.  Tabulated approaches also provide an effective control on the thermodynamic error of the state equation. For single species fluid,  two independent thermodynamic variables are needed to fully define the local thermodynamic state in a single phase.  Classically, the choice of the thermodynamic tuple has been pressure and temperature given the convenience of computing the isothermal and isobaric conditions.  For CFD simulations, this approach is sub-optimal since the known transported quantities are $\rho$ and $e$; thus making these the ideal basis for any thermodynamic table. Once the corresponding position in the table is found, all thermodynamic and transport properties may be interpolated or computed.

\begin{figure}[h]
\centering
\hspace*{-0.5cm}
\includegraphics[scale= 0.43]{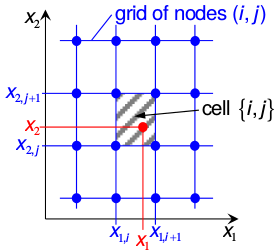}
\caption{A typical cell in a table, the red dot indicating the unknown point \label{fig:lookup}}
\end{figure}

 In this work, we consider three different tabulation approaches:
 \begin{itemize}
 \item Homogeneous of logarithmic tabulation
 \item Block structured Adaptive Mesh Refinement (AMR)
 \item Adaptive Mesh Refinement with Bezier patch fitting 
 \end{itemize}
Details of each tabulation approach is presented in the following:

\subsection{Homogeneous or logarithmic tabulation}
\begin{figure}[h]
\centering
\hspace*{-0.5cm}
\includegraphics[scale= 0.43]{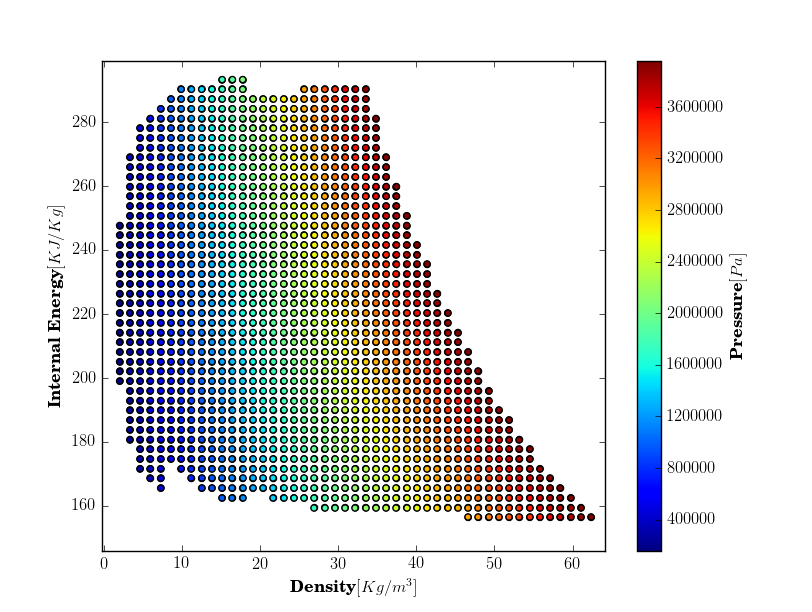}
\caption{Uniformly spaced lookup table for oxygen \label{fig:homogeneous}}
\end{figure}

The simplest data structure for tabular reconstruction is one that spans the region of thermodynamic interest with equally spaced grids, or alternatively under a known mapping function such as a logarithmic distribution. Equal spacing provides very efficient table look up because the reconstruction interval can be computed directly without searching, but this results in very large table sizes as the grid size is not adapted to the magnitude of the local error. The level of accuracy is set by the region in which the function varies most rapidly so that slowly varying regions must be over-refined to accommodate high-gradient regions. Tabulating in density/energy space, as shown in figure \ref{fig:homogeneous}, allows for a simple lookup from the transport equations but simultaneously the corner regions of the table contain pressure and temperature values that are far outside the region of interest for most CFD problems. Therefore, bounds typical isobaric conditions of the simulation are mapped to the energy/density space. 

\subsection{Block Structured Adaptive Mesh Refinement (AMR)}
The adaptive mesh refinement table is based on Cartesian adaptive mesh structure.   We begin by defining a rectangular region in the x-y plane (P-T plane for this example) that includes the physical domain of interest. The required properties (internal energy, density, enthalpy, entropy, etc.) are evaluated at each of the four corners of the rectangle and appropriate reconstruction is used to evaluate these properties at pre-selected points in the rectangle. For simplicity, we chose bivariate interpolation to evaluate the properties at those preselected 25 equidistant points. The reconstructed (interpolated) properties are compared with the 'exact' values obtained from the NIST database to assess the reconstruction (interpolation) error. If the error is more than the user-specified tolerance, it is subdivided into four equal rectangles with property evaluations again being computed and compared with the exact NIST database values. As the interpolation grid is refined, relatively flatter(linear) portions of the surface will fall within the error tolerance and will not require further refinement while rectangles in the regions of larger gradients will be refined further. Keeping track of these adaptively refined regions in a tree-based structure enables rapid search procedures that can be comparable to the speed of equally of spaced tables.

\begin{figure}[H]
\centering
\hspace*{-0.7 cm}
\includegraphics[scale= 0.13]{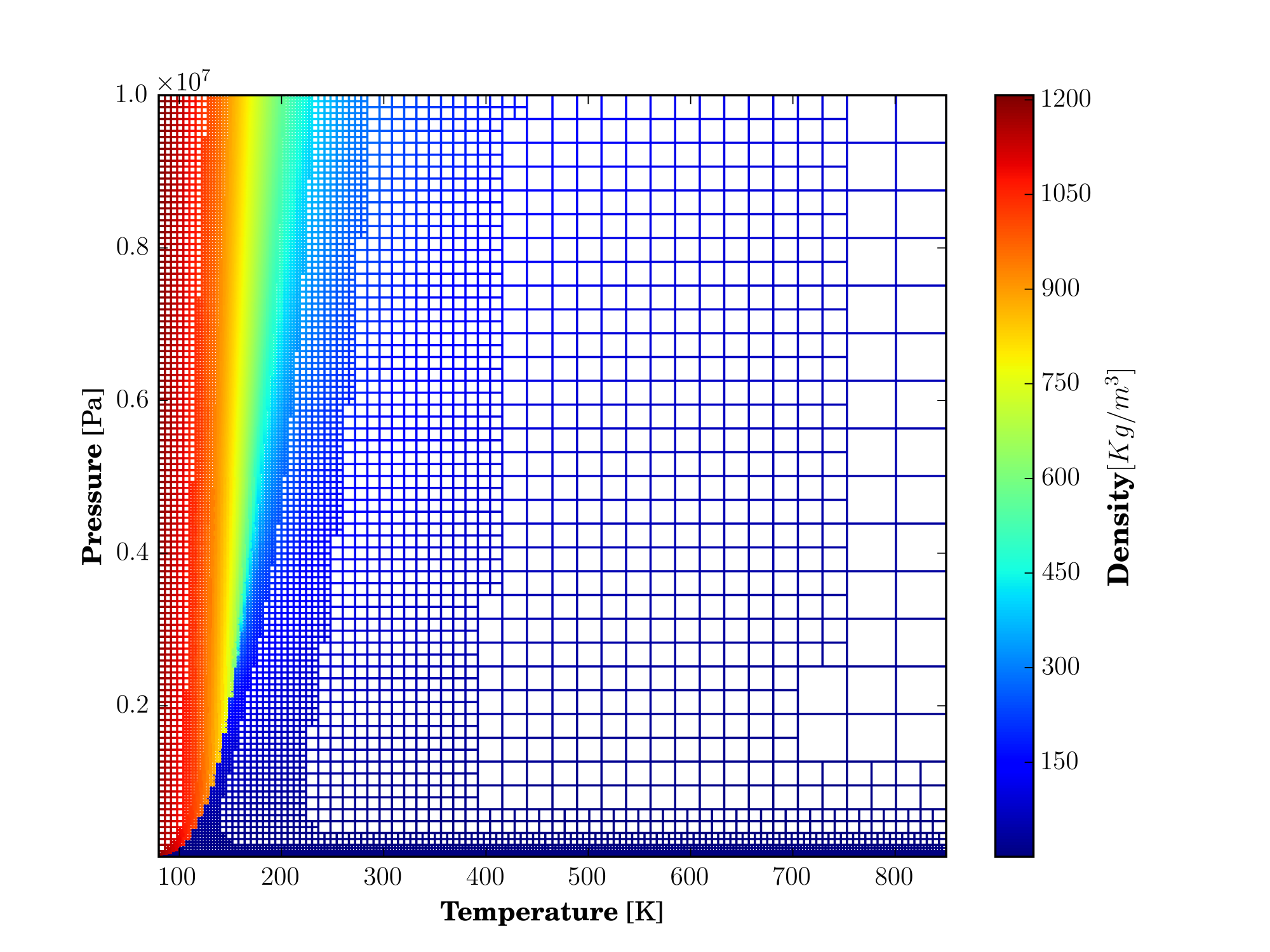}
\caption{Block structured AMR lookup table for oxygen with 0.01\% accuracy }
\end{figure}

\subsubsection{Lookup Approach}
The Structured block based adaptive mesh refinement has been widely used in problems with highly oscillatory phenomenon, as it provides uniform interpolation accuracy over the entire thermodynamic region of interest.  The patch (in this case, the rectangular cell) must be located before bilinear interpolation is used to reconstruct the thermodynamic property. One easy straightforward method is to search the patches one by one until we zero-in onto the last child node (patch) that contains the given unknown coordinates (Input thermodynamic parameters T,P or $\rho$, E). This sequential searching algorithm's efficiency strongly depends on the table size thereby inhibits speed if the thermodynamic domain of interest is large and the depth of refinement is high. 

\begin{figure}[H]
\centering
\hspace*{-0.7 cm}
\includegraphics[scale= 0.35]{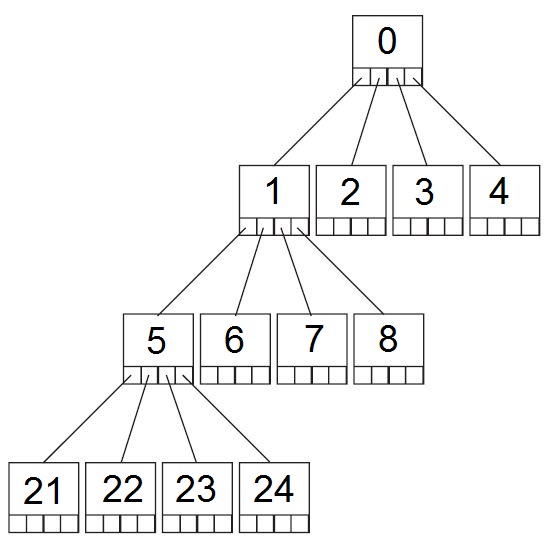}
\caption{Sample quadtree subdivision }
\end{figure}

As an alternative approach, we employ a fast lookup algorithm that requires additional memory to store the indices in tree data-structure, and the efficiency doesn't depend much on the depth of the tree. It is constructed by the following steps, 

A uniform index array based on quadtree subdivision covering the entire computational domain is generated as per the maximum refinement level of the generated adaptive table. For example, at a given maximum refinement level of 3, the uniform index table is shown in the figure 4. If the maximum refinement level is n+1 (the refinement level of the single rectangular cell that cover the entire thermodynamic region of interest is 0) then there will be $4^n$ number of individual patches (in this case rectangular cells), the integer number stored in the cell indicates the index of that patch.

An overlapping mapped index array is also generated which contains the child node index in the place of all further sub divisions pertaining to that child if there was a refinement as illustrated in figure 5.
The cell within which a point $(T_x, P_x)$ is located can be determined at an algorithmic complexity $O(1)$, directly without searching the tree (which is of algortimic complexity $O(n)$ where n being the refinement level) , by using the mapping index array. Furthermore, the storage required for the uniform index table is small for this generation computers, for example, if the maximum refinement level of the adaptive table is 10, and an integer requires 4 B of memory, the uniform index table requires $(2^{9}x2^{9}x4)/(1024^2) MB = 1 MB$.

%%-------------Needs to be regenerated with smaller refinement size---%%%
% \begin{figure}[H]
% \centering
% \hspace*{-0.8 cm}
% \includegraphics[scale= 0.35]{uniform_index}
% \caption{Sample uniform index for a tree depth of 4}
% \end{figure}

\begin{figure}[H]
\centering
\begin{tikzpicture}[draw=black, thick, x=\Size,y=\Size]
    \foreach \col/\colLetter in \Sequence {%
        \foreach \row/\rowLetter in \Sequence{%
            \pgfmathtruncatemacro{\value}{\col+\NumOfColumns*(\row-1)}
            \def\NodeText{\expandafter\csname Node\rowLetter\colLetter\endcsname}
            \node [Square] at ($(\col,-\row)-(0.5,0.5)$) {\NodeText};
        }
    }
\end{tikzpicture}
\caption{Uniform index for a maximum refinement level of 3}
\end{figure}

\begin{figure}[H]
\centering
\hspace*{-0.7 cm}
\includegraphics[scale= 0.07]{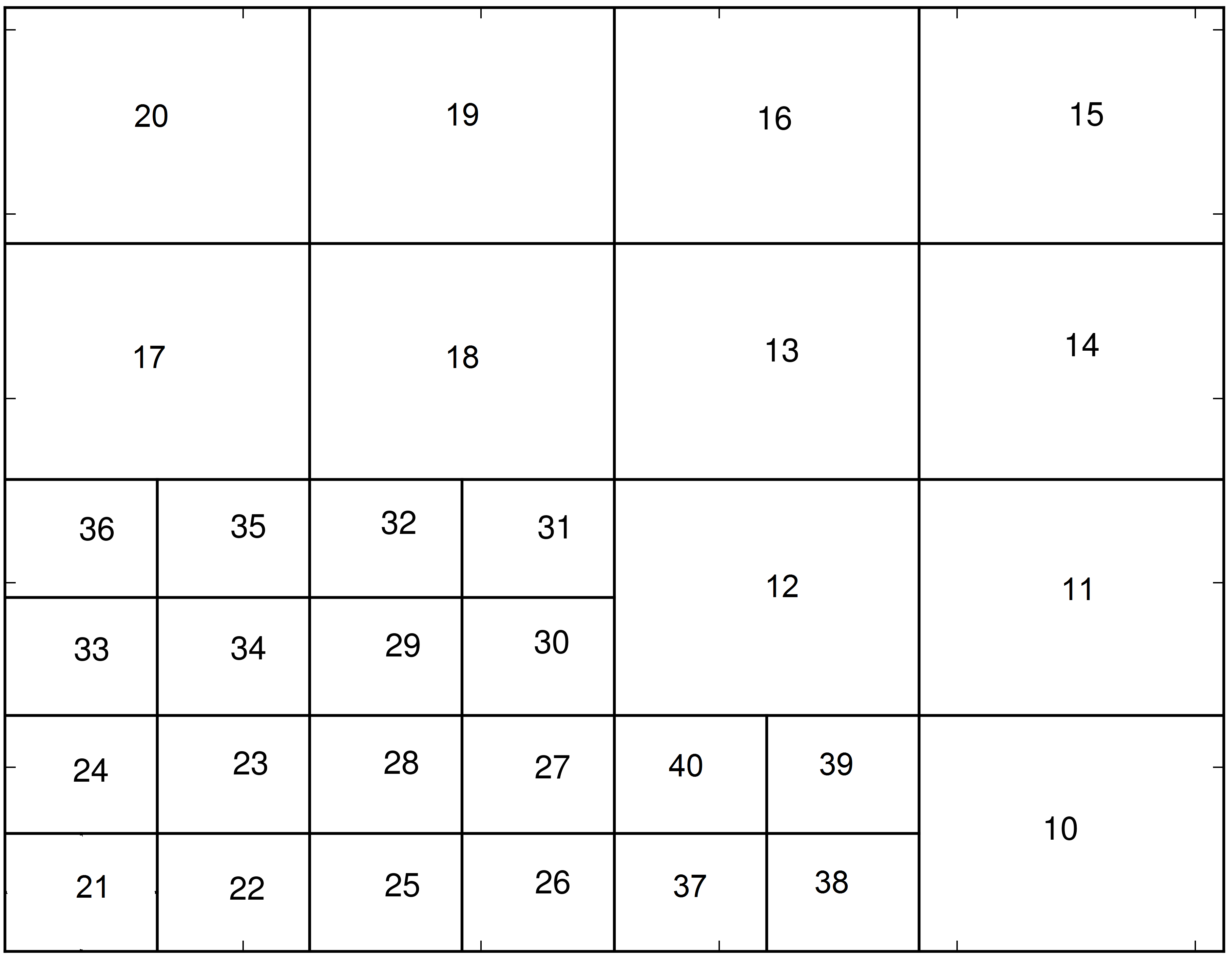}
\caption{Sample Overlapping index for the uniform index at the maximum refinement level = 3 }
\end{figure}

\subsection{Bezier patch with AMR}
In the block structured AMR table, the underlying criteria for refinement is based on the bivariate approximation. Rather than tabulating solely on the basis of bilinear reconstruction error we chose to implement Bezier patches that are approximated surfaces. The approximated function of two independent variables is given by \cite{Collins2013} 

\begin{equation}
%F(u, v) = {\mathlarger{\sum}}_{n=0}^{\infty}
F(u, v) = \mathlarger{\mathlarger{\sum}}_{i=0}^{3}\mathlarger{\mathlarger{\sum}}_{j=0}^{3}B_i^3 (u)B_j^3(v)b_{ij}
\end{equation}

\begin{figure}[H]
\centering
\hspace*{-0.85 cm}
\includegraphics[scale= 0.39]{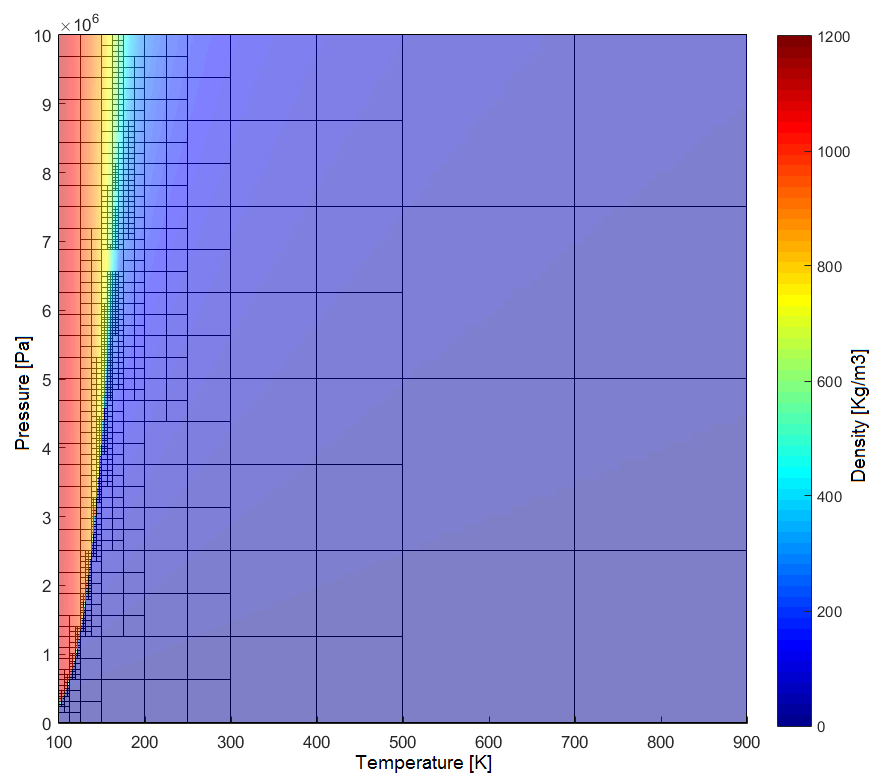}
\caption{Bezier patch based Tabulated EoS for Density($\rho$) adaptively refined using Quadtree}
\end{figure}

Here, $B_i^3(u)$ are the third-degree (fourth-order) Bernstein polynomials, $b_{ij}$ are the control points. 
\begin{equation}
B_i^3(u) = \frac{3!}{i!(3-i)!}(1-u)^{3-i}u^i
\end{equation}
Using the linear combination of these $b_{ij}$ control points, the shape of the surface is determined. The surface is generated by recursively subdividing the domain until the resulting Bezier patches are capable of reconstructing the original value to a specified error threshold or if the maximum refinement level is reached.

\section{Uncertainty quantification of Tabular equation of State}

An open problem in tabular equation of states is uncertainty visualization and quantification, how uncertainty and interpolation interact. There are many established interpolation schemes available, but how reliable are the interpolated values ? We want to address this problem by analyzing the interpolation in the tabular equation of state and its impact on gaussian distribution data. 

An important feature of the tabular representation of Equation of state is the feasibility to represent the underlying uncertainty associated within the thermodynamic properties. This helps in better understanding of the uncertainty propagation through the CFD code.

We followed this framework for parametric uncertainty quantification in analytical equation of state.

\begin{itemize}
	\item Identify and characterize all sources of input uncertainty, aleatory and epistemic.
    \item Propagate input uncertainties through the computation.
    \item Estimate uncertainty due to numerical approximation.
    \item Determine total uncertainty in the predicted quality of uncertainty.
\end{itemize}

Inorder to quantify the uncertainty, we should begin with the underlying analytical form of the equation of state with which the points are tabulated.
analytical equation as in a mathematical form of equation of state is an approximate representation of an element's complex thermodynamic system. Such mathematical forms (models) use a large number of model parameters ,whose exact values can't be determined analytically most of the times, resulting in the fitting of model predictions with calibration data obtained from experiments or much more highly defined models. The author has chosen Peng-Robinson equation of state to demonstrate the UQ, as the highly accurate Benedict-Webb-Rubin(BWR) equation has more than 11 parametric constraints it would be intensive.

\begin{eqnarray}
P &=& \frac{RT}{V_m - b} - \frac{a\alpha}{{V_m}^2 + 2bV_m - b^2} \\
a &=& \frac{0.45724R^2{T_c}^2}{p_c} \nonumber \\
b &=& \frac{0.07780RT_c}{p_c} \nonumber \\
\alpha &=& (1+\kappa(1 - {\frac{T}{T_c}}^{0.5}))^2 \nonumber \\
\kappa &=& 0.37464 + 1.54226\omega - 0.26992\omega^2 \nonumber
\end{eqnarray}

For the Peng Robinson equation of state the uncertain model parameters are obtained from \cite{rowley2007database}

\begin{table}[H]
\centering
\caption{Parametric Uncertainty in Peng Robinson Equation of state}
\label{my-label}
\begin{tabular}{|l|l|}
\hline
Parameter & Uncertainty(\%) \\ \hline
Critical Temperature* {[}T{]} & \rpm 3 \\ \hline
Critical Pressure* {[}p{]} & \rpm 5 \\ \hline
Accentric factor {[}$\omega${]} & \rpm5 \\ \hline
$k^{\#}$ & \rpm5 \\ \hline
\end{tabular}
    \begin{tablenotes}
      \small
      \item \-*Experimental, \#-Assumed
    \end{tablenotes}
\end{table}

%\subsubsection{UQ in analytical equation of states} 

% \begin{table}[H]
% \centering
% \caption{Parametric Uncertainty in Peng Robinson Equation of state}
% \label{my-label}
% \begin{tabular}{|l|l|l|l|l|}
% \hline
% \begin{tabular}[c]{@{}l@{}}Para- \\ meter \end{tabular}  & \begin{tabular}[c]{@{}l@{}}Uncert-\\ ainty \\ (\%)\end{tabular} & \begin{tabular}[c]{@{}l@{}}Lower \\ bound\end{tabular} & \begin{tabular}[c]{@{}l@{}}Nom- \\ inal\end{tabular} & \begin{tabular}[c]{@{}l@{}}Upper\\  bound\end{tabular} \\ \hline
% \begin{tabular}[c]{@{}l@{}}Critical \\ Temp.[K] \end{tabular} & \rpm 3 & 238.232 & 245.60 & 252.968 \\ \hline
% \begin{tabular}[c]{@{}l@{}}Critical \\ Pressure \end{tabular} & \rpm 5 & 18.424 & 19.394 & 20.363 \\ \hline
% \begin{tabular}[c]{@{}l@{}}Accentric \\ Factor$\omega$\end{tabular} & \rpm 5 & 0.398 & 0.419 & 0.440 \\ \hline
% k & \rpm 5 & -0.055 & -0.053 & -0.050 \\ \hline
% \end{tabular}
% \end{table}

\begin{figure}[H]
\centering
\hspace*{-0.5 cm}
\includegraphics[scale = 1]{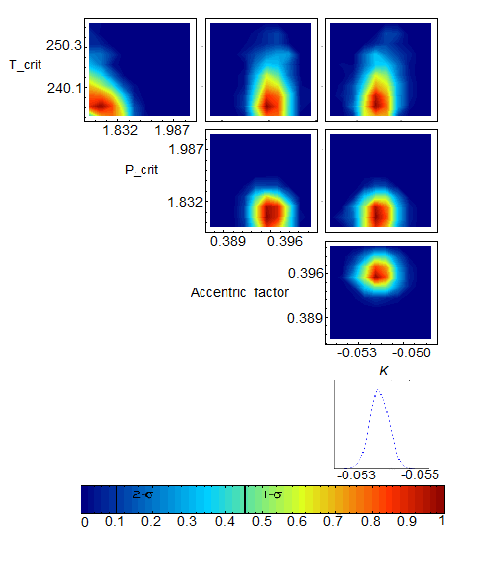}
\caption{The contour plots show the two-dimensional distributions of all pairs of parameters, where the color scale represents the likelihood value normalized to the maximum likelihood. The levels corresponding to $1-\sigma$ and $2-\sigma$ regions are indicated in the color bar.}
\end{figure}

\begin{figure}[H]
\centering
\hspace*{-0.5 cm}
\includegraphics[scale = 0.42]{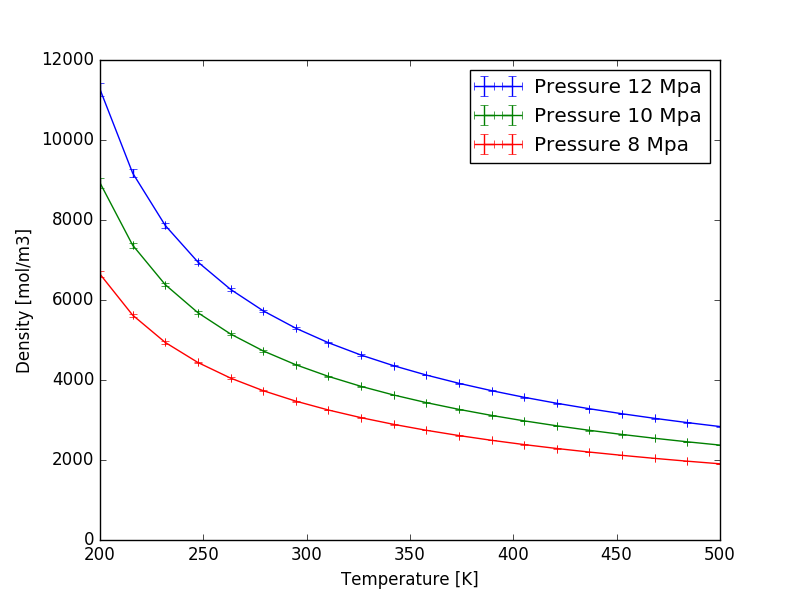}
\caption{Density computed with UQ enabled Peng Robinson Equation of state for Oxygen}
\end{figure}
In this paper, one of the objective is to propagate this model parameter uncertainty through the flux calculations for the solver to understand the statistical behaviour of output properties.

%LHS (Latin Hypercube sampling)

\subsection{UQ in Tabular equation of state}

We analyzed the effect of bilinear interpolation of a data with four uncorrelated scalar Gaussian distributions $X_i \sim N(\mu_i, {\sigma_i}^2)$ at the positions $s_i$, where i = 1,2,3,4.

\begin{figure}[H]
\centering
%\hspace*{-0.3 cm}
\includegraphics[scale= 0.25]{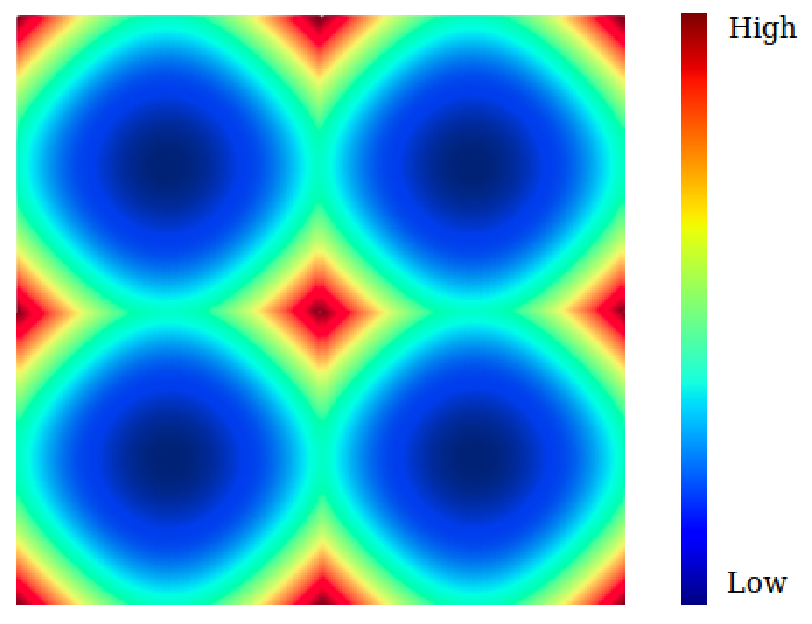}
\caption{Color coded variance as the result of Bilinear interpolation of uncorrelated
Gaussian distributed variables given on the grid points}
\end{figure}

By our understanding, bilinear interpolation is in general not well suited to estimate uncertain values. A key element of interpolation is that the interpolation result equals the value at the data points. In the case that the value is uncertain, this property may be unwanted because of that data point is disregarded. And the result of the variance might be that the best estimator for that data point is actually not the give value at that data point.

We point out that the traditional bilinear interpolation, which is widely used in tabular equation of state may yield very counter-intuitive results when dealing with uncertain data. If the covariance between neighbouring points is weaker than the variance at these points, then the resulting interpolated data is most certain between the grid points (where we have no data) and the variance has its maxima at the grid points (where the data is given). If one encounters such data, it should be obvious by now that one needs other methods.

\subsection{Thermodynamic consistency}
Thermodynamic consistency in property evaluations has been discussed by Swesty\cite{Swesty1996}, using the Helmholtz free energy as a basis function.
The problem of thermodynamic consistency is very important in many cases where reactive flows need to be simulated over very long timescales.
Thermodynamic inconsistency may become manifest via the unphysical buildup of entropy, or temperature.

Thermodynamic consistency requires that the first law of thermodynamics be satisfied, so by referring to how well the EOS satisfies the constraints posed by Maxwell relations we can quantify the thermodynamic consistency\cite{PhysRevE.73.066704}.

If we have internal energy E expressed as a function of specific volume V and entropy S. The thermodynamic definitions of pressure P and temperature T are

\begin{equation}
        P = -\frac{\partial E}{\partial V}\Bigg|_S,  T = -\frac{\partial E}{\partial S}\Bigg|_V
\end{equation}

With the bilinear interpolation, both the density and its first order derivatives vary linearly over each local rectangle. Further, enthalpy and density have no interdependence on each other. Nevertheless, as the thermodynamic properties stored at the mesh nodes are themselves taken from a consistent thermodynamic database (in this case NIST), the internal inconsistencies will be small on a fine grid.

\subsection{Error estimates and Verification}
The T-P tabular grid is sampled into 1000x1000 log-linear grid. In each patch 100 points are sampled and then the maximum relative error in density is computed. As seen from the relative error contour plot, the maximum error is along the phase change line or where the density changes rapidly. Except for that region the uniform accuracy is maintained as a result of adaptive subdivision. The highest error points lies close to the critical point. The number of patches required to capture the thermodynamics is different in three approaches (Homogeneous, Adaptive, Bezier patch) The highest being the homogeneous due to its uniform spacing, it will take enormous number of patches to get the error close to 0.01\%, althought Bezier patch has less number of patches it can only accommodate one property at a time, which means it requires individual table for each property. The Block based AMR has moderate number of patches that are not oversized but can tabulate all thermodynamic properties.

\begin{figure}[H]
\centering
\hspace*{-0.7 cm}
\includegraphics[scale= 0.44]{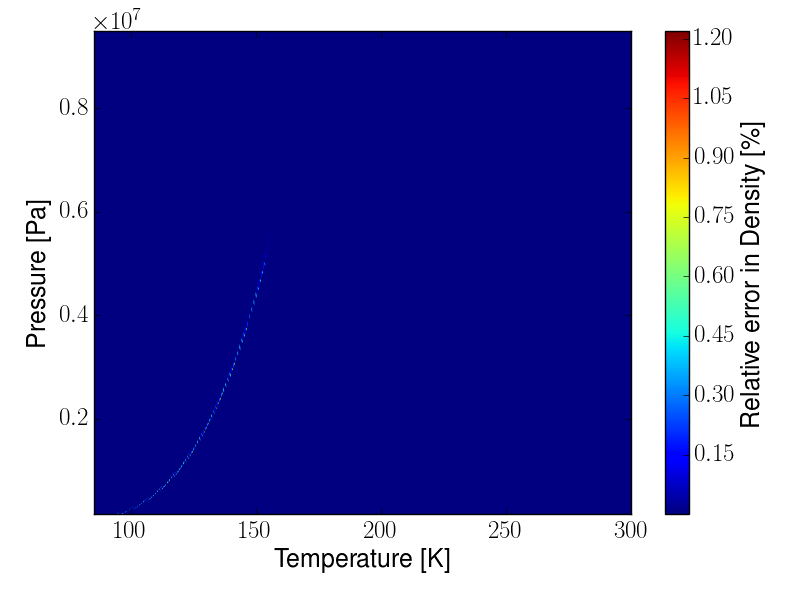}
\caption{Relative error in density for the Block based adaptively refined table}
\end{figure}

\begin{table}[H]
\centering
\caption{Error estimates}%\tablefootnote{Compared to NIST} }
\label{my-label}
\begin{tabular}{|l|l|l|}
\hline
Grid type & \begin{tabular}[c]{@{}l@{}}Block based AMR\\ (0.0001 tol.)\end{tabular} & \begin{tabular}[c]{@{}l@{}}Bezier Patch \\ AMR\\ (0.0001 tol.)\end{tabular} \\ \hline
\begin{tabular}[c]{@{}l@{}}Total \\ patches\end{tabular} & 12648 & 1120 \\ \hline
\begin{tabular}[c]{@{}l@{}}\% patches\\  with rho\\  error \textgreater 0.001\end{tabular} &  0.101 & - \\ \hline
\begin{tabular}[c]{@{}l@{}}Maximum\\  rho error  $\%$ \end{tabular}  &  1.2 & - \\ \hline

\end{tabular}
\end{table}

\begin{figure}[H]
\centering
\hspace*{-0.7 cm}
\includegraphics[scale= 0.44]{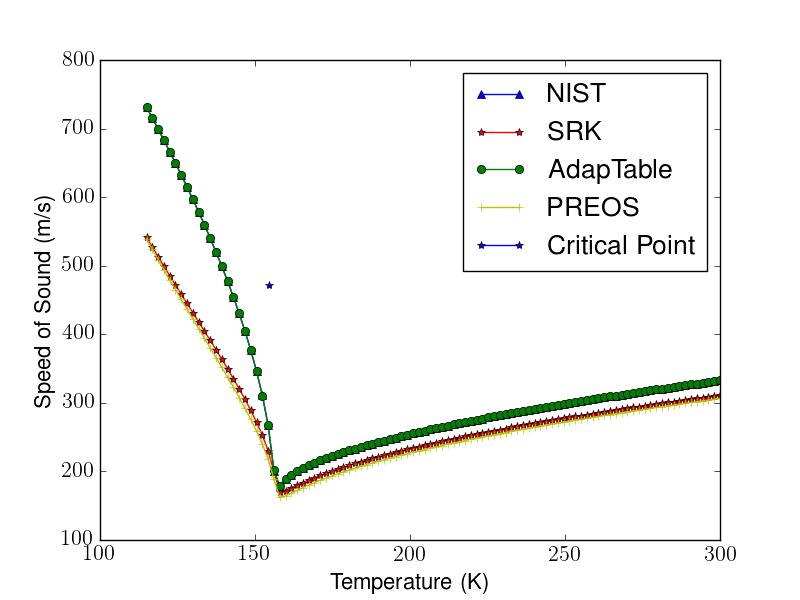}
\caption{Error plot for speed of sound for Oxygen computed with different equation of states}
\end{figure}

The comparison of various tabulation approaches demands a quantitative thermodynamic error metric.  As the exactness thermodynamic properties remain a subject of scientific  relevance, the current study evaluates the error relative to the NIST database. These data points represent the  highest fidelity data set for the computation of the thermodynamic and thermophysical properties and constructed  from an ensemble of experimental data supplemented with  Modified Benedict-Webb-Rubin curve fitting. %We propose three metrics to estimates to thermodynamic error that will be detailed in the final paper:

To evaluate different types of equation of states, statistical parameters are used, 
The Average Percent Relative Error (ARE\%), which is defined below is a measure of the bias of the correlation; a value of zero indicates a random of the measured values around the correlation.
\begin{eqnarray}
ARE\% = \frac{100}{N_d} \mathlarger{\mathlarger{\sum}}_{i = 1}^{N_d}\frac{(\rho_i^{exp.} - \rho_i^{calc.})}{\rho_i^{exp.}}
\end{eqnarray}
The AARE\%, is the arithmetic average of the absolute values of the relative errors; is an indication of the accuracy of the correlation.
\begin{eqnarray}
AARE\% = \frac{100}{N_d} \mathlarger{\mathlarger{\sum}}_{i = 1}^{N_d}\frac{|\rho_i^{exp.} - \rho_i^{calc.}|}{\rho_i^{exp.}}
\end{eqnarray}
The $R^2$, is the correlation coefficient; is a measure of the precision of fit of the data. If data are perfectly correlated, then $R^2 = 1$. A small value of AARE\% and $R^2$ close value to one (simultaneously) denote a good correlation based on good data.

Another parameter, Sum of Absolute of Residual (SAR) shows the reliability of correlation for higher order data points. 
\begin{eqnarray}
SAR\% = \mathlarger{\mathlarger{\sum}}_{i = 1}^{N_d}|\rho_i^{exp.} - \rho_i^{calc.}|
\end{eqnarray}

The statistical parameters of the well known EoSs are listed in table 1.

\begin{table}[H]
\centering
\caption{Error metrics comparison with NIST}
\label{my-label}
\begin{tabular}{|l|l|l|l|l|}
\hline
\begin{tabular}[c]{@{}l@{}}Equation\\  of State\end{tabular} & AARE & ARE & \begin{tabular}[c]{@{}l@{}}SAR \\ {[}Kg/m\textasciicircum 3{]}\end{tabular} & R\textasciicircum 2 \\ \hline
\begin{tabular}[c]{@{}l@{}}Peng-\\ Robinson\end{tabular} & 3.015 & 0.993 & 70.600 & 0.991 \\ \hline
\begin{tabular}[c]{@{}l@{}}Redlich-\\ Kwong\end{tabular} & 4.409 & -0.026 & 89.816 & 0.987 \\ \hline
\begin{tabular}[c]{@{}l@{}}Adap-\\ Table\end{tabular} &  1.022 & 0.0021 & 91.232  & 0.995 \\ \hline
\begin{tabular}[c]{@{}l@{}}Soave-\\ Redlich-\\ Kwong\end{tabular} & 6.388 & 6.381 & 131.009 & 0.990 \\ \hline
\end{tabular}
\end{table}

\section{Computational cost estimates}
The increase in speed of evaluating properties compared to cubic EOS is one of the primary motivations. Here we show an example of the speedup of computations. When pressure is evaluated from randomly generated $\rho$, $e$ combination in the supercritical regime. For each EOS, 100000 loops/calls are executed and the slowest run time is tabulated.

Searching algorithm and interpolation scheme influence the efficiency of the table tabular equation of state, the adaptive mesh is comparably denser in the near critical point regime, because of high non linearity which resulted in comparatively higher lookup search times for those regions as shown in the figure 13

\begin{figure}[h]
\centering
\hspace*{-0.66 cm}
\includegraphics[scale = 0.42]{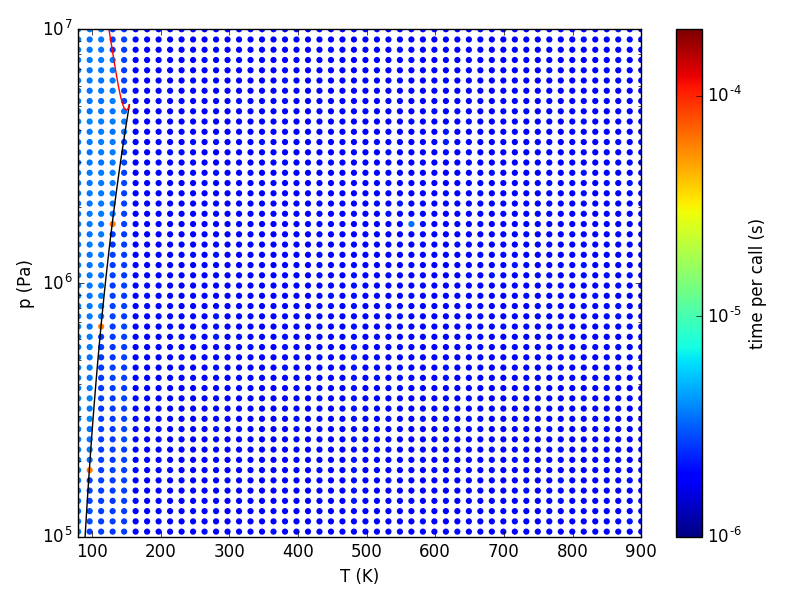}
\caption{Block based tabulated state equation lookup Computational Cost in $\mu$ sec }
\end{figure}

\begin{figure}[h]
\centering
\hspace*{-0.66 cm}
\includegraphics[scale = 0.42]{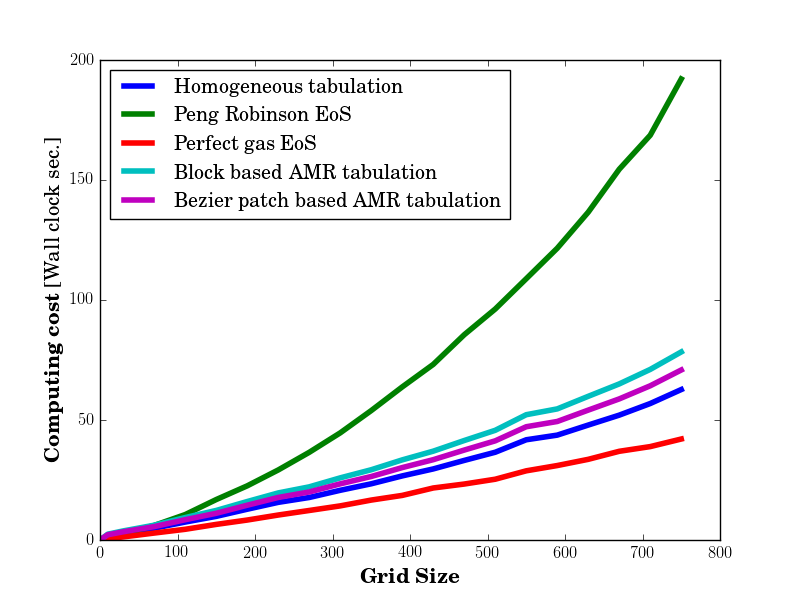}
\caption{Travelling Acoustic wave, one flow-through time period simulation computational cost comparison with different equation of states}
\end{figure}
% \begin{tabular}{ |p{3cm}||p{3cm}|p{2cm}|p{3.5cm}|  }
%  \hline
%  \multicolumn{4}{|c|}{Computation cost ($\mu$ sec) per call} \\
%  \hline
%  Element  & Peng\_Robinson EOS &SRK EOS& Adaptive lookup table(0.1\%)\\
%  \hline
%  Propane   & 19.86    &20.9&   14.6\\
% Oxygen&   17.64  &    19.11& 14.35\\
%  Carbon dioxide &11.36 & 12.72&  8.36\\
%  \hline
% \end{tabular}

\section{One dimensional test case:}
A one dimensional test case is identified to evaluate the performance of the proposed tabulated equation of state.

% \section*{Diffusion}:
% \begin{figure}[h]
% \centering
% %\hspace*{-2.5cm}
% \includegraphics[scale = 0.25]{diff_p.png}
% \caption{Velocity distribution in Nozzle \label{fig:homogeneous}}
% \end{figure}

% \section*{Converging - Diverging Nozzle}:
% A converging - Diverging Nozzle of throat to Exit area ratio of 0.5 is considered, for the inlet pressure of 101325 Pa, the calculated exit pressure is 96258.75 Pa. Dirichlet boundary conditions are applied at the inlet and exit. 
% \begin{figure}[H]
% \centering
% %\hspace*{-2.5cm}
% \includegraphics[scale = 0.25]{nozzle_p.png}
% \caption{Pressure distribution in Nozzle }
% \end{figure}

% \begin{figure}[H]
% \centering
% %\hspace*{-2.5cm}
% \includegraphics[scale = 0.25]{mach_nozzle.png}
% \caption{Velocity distribution in Nozzle }
% \end{figure}

\section*{Harmonic Acoustic wave}:
In one-dimensional acoustic wave propagation in supercritical fluid case, Using periodic boundary conditions at both sides a harmonic wave is initialized. The computational domain is $x \in {[0, 10]}$ m and 100 grid points are used, giving a uniform grid spacing $\Delta x = 0.1m$. 
The sound pressure level $L_p = 23$ (dB) is calculated according to:
\begin{equation}
L_p = 20log_{10}(\frac{\Delta p}{p_{ref}\sqrt{2}})
\end{equation}
Where $\Delta p$ is the amplitude of the pressure harmonic wave (Pa).

As shown in the figure, there is a significant effect in the form of density by using most accurate tabular real fluid thermodynamics (AdapTable) rather than conventional Peng robinson equation of state (PREOS)

\begin{figure}[H]
\centering
\hspace*{-1.1 cm}
\includegraphics[scale = 0.30]{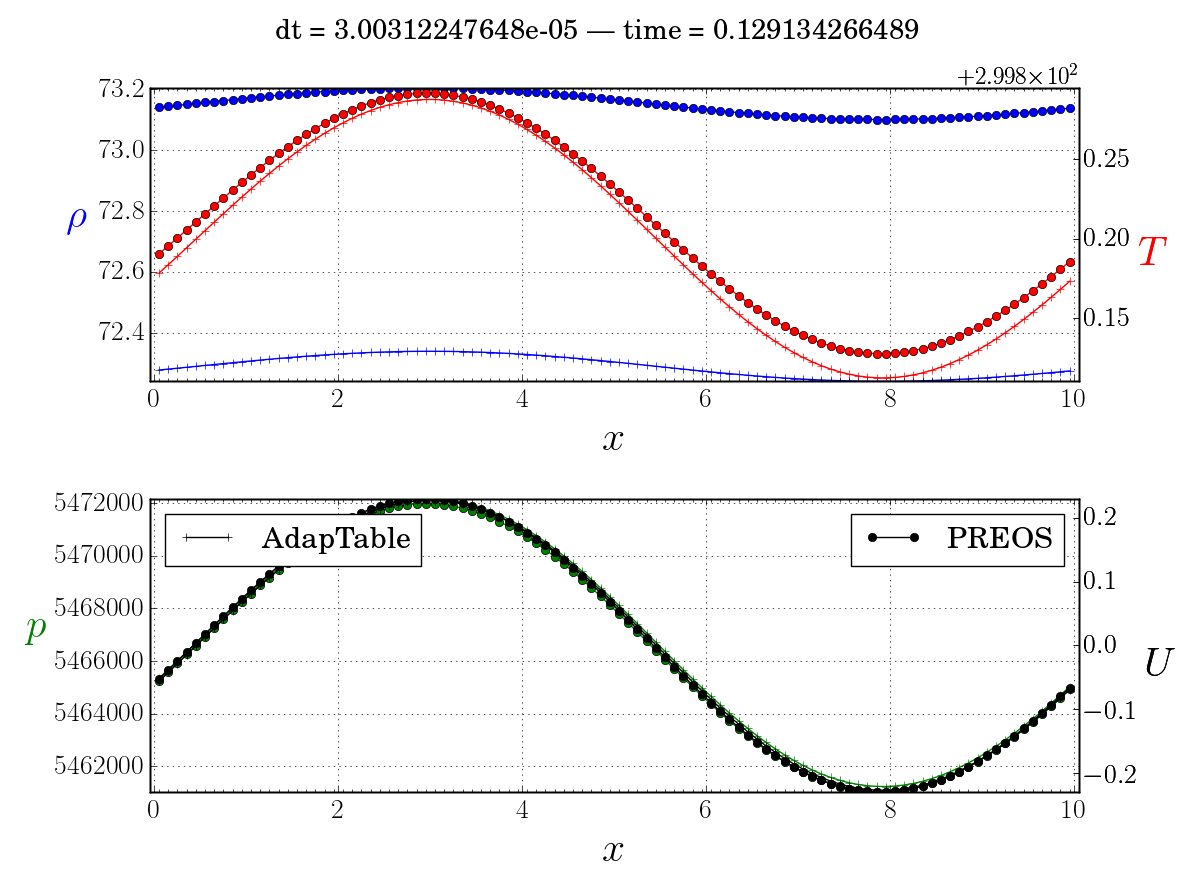}
\caption{Harmonic Acoustic wave propagation }
\end{figure}

\section*{Conclusion}

We have demonstrated three different types of tabulated equation of states for an efficient calculation of thermodynamic properties of real and supercritical fluids. The EOS look-up tables are constructed using Block based AMR, Bezier patch based AMR, homogeneous tabulation using bilinear interpolation. Their error metrics, computational cost and uncertainty in such tabular representation of gaussian distributed data is discussed along with a test case highlighting the use of a tabular real gas equation of state in a CFD code.

\bibliographystyle{unsrt}
\bibliography{journaltempnotes}

\begin{thebibliography}{10}

\bibitem{Aizpurua-Olaizola2015}
O.~Aizpurua-Olaizola, M.~Ormazabal, A.~Vallejo, M.~Olivares, P.~Navarro,
  N.~Etxebarria, and A.~Usobiaga.
\newblock Optimization of supercritical fluid consecutive extractions of fatty
  acids and polyphenols from vitis vinifera grape wastes.
\newblock {\em Journal of Food Science}, 80(1):E101--E107, 2015.

\bibitem{Cai2014}
J.~Cai, C.~Renault, and J.~Gou.
\newblock Supercritical water-cooled reactors.
\newblock {\em Science and Technology of Nuclear Installations}, 2014.

\bibitem{Cinnella2011}
P.~Cinnella, P.~M. Congedo, V.~Pediroda, and L.~Parussini.
\newblock Sensitivity analysis of dense gas flow simulations to thermodynamic
  uncertainties.
\newblock {\em Phys. Fluids}, 2011.

\bibitem{Collins2013}
E.~Collins and E.~Luke.
\newblock Fast evaluation of complex equations of state.
\newblock {\em Electronic Journal of Differential Equations}, 2013.

\bibitem{PhysRevE.73.066704}
G.~A. Dilts.
\newblock Consistent thermodynamic derivative estimates for tabular equations
  of state.
\newblock {\em Phys. Rev. E}, 73:066704, Jun 2006.

\bibitem{Hickey2013}
J.-P. Hickey and M.~Ihme.
\newblock {\em Center for Turbulence Research - Annual Research Briefs 2014},
  chapter Supercritical mixing and combustion in rocket propulsion, pages
  21--36.
\newblock Stanford University, 2013.

\bibitem{Liu2014}
Z.~Liu, J.~Liang, and Y.~Pan.
\newblock Construction of thermodynamic properties look-up table with
  block-structured adaptive mesh refinement method.
\newblock {\em J. Thermophysics. Heat Transfer}, 2014.

\bibitem{Liu2015}
Z.~Liu, J.~Liang, and Y.~Pan.
\newblock Efficient thermodynamic properties reconstruction method with
  adaptive triangular mesh.
\newblock {\em J Thermophys Heat Transfer}, 2015.

\bibitem{Masquelet2006}
M.~M. Masquelet.
\newblock {\em Simulations of a sub-scale liquid rocket engine: Transient heat
  transfer in a real gas environment}.
\newblock PhD thesis, Georgia Institute of Technology, 2006.

\bibitem{rowley2007database}
R.~L. Rowley, W.~V. Wilding, J.~L. Oscarson, and Y.~Yang.
\newblock Database tools for evaluating thermophysical property data.
\newblock {\em International Journal of Thermophysics}, 28(3):805--823, 2007.

\bibitem{Stryjek1986}
R.~Stryjek and J.~Vera.
\newblock Prsv: An improved peng-robinson equation of state for pure compounds
  and mixtures.
\newblock {\em The canadian journal of chemical engineering}, 64(2):323--333,
  1986.

\bibitem{Stucki2009}
S.~Stucki, F.~Vogel, C.~Ludwig, A.~G. Haiduc, and M.~Brandenberger.
\newblock Catalytic gasification of algae in supercritical water for biofuel
  production and carbon capture.
\newblock {\em Energy \& Environmental Science}, 2(5):535--541, 2009.

\bibitem{Swesty1996}
F.~D. Swesty.
\newblock Thermodynamically consistent interpolation for equation of state
  tables.
\newblock {\em Journal of Computational Physics}, 127(1):118--127, 1996.

\end{thebibliography}

\end{document}